\documentclass[12pt,a4paper]{article}
\usepackage{epsf}
\usepackage{cite}
\usepackage{amsmath,amssymb}
\usepackage[top=2.5cm, bottom=3cm, left=2.5cm, right=2.5cm]{geometry}
\usepackage[dvipdfmx]{graphicx, xcolor}
\usepackage{comment}
\usepackage{ulem}

\begin{document}

\newcommand{\rem}[1]{{\textcolor{red}{$\spadesuit$\bf #1$\spadesuit$}}}

\begin{titlepage}

\begin{center}

\hfill \\
\hfill UT-15-03\\

\vskip 1.5in

{
\large \bf Gravitational Effects on Inflaton Decay
}

\vskip .75in

{
Yohei Ema$^a$, Ryusuke Jinno$^a$, Kyohei Mukaida$^a$, Kazunori Nakayama$^{a,b}$
}

\vskip 0.25in

\vskip 0.25in

{\it $^a$Department of Physics, University of Tokyo, Tokyo 113-0033, Japan}\\
\vskip 0.1cm
{\it $^b$Kavli IPMU, TODIAS, University of Tokyo, Kashiwa 277-8583, Japan}

\end{center}
\vskip .5in

\begin{abstract}

We point out that the inflaton inevitably couples to all non-conformally coupled matters gravitationally through 
an oscillation in the Hubble parameter or the cosmic scale factor.
It leads to particle production during the inflaton oscillation regime, which is most efficient just after inflation.
Moreover, the analysis is extended to the model with non-minimal inflaton couplings to gravity,
in which the Hubble parameter oscillates more violently.
We apply our results to the graviton production by the inflaton: gravitons are also produced just after inflation,
but the non-minimal coupling does not induce inflaton decay into the graviton pair.

\end{abstract}

\end{titlepage}

\renewcommand{\thepage}{\arabic{page}}
\setcounter{page}{1}
\renewcommand{\thefootnote}{$\spadesuit$\arabic{footnote}}
\setcounter{footnote}{0}
\renewcommand{\theequation}{\thesection.\arabic{equation}}

\section{Introduction}
\setcounter{equation}{0}

Inflation explains both the homogeneous and isotropic universe and the small density fluctuation in a beautiful way.
In a successful inflation model, the energy density of the inflaton must be converted into high-temperature radiation.
This process is called reheating.

In a standard picture, the reheating is caused by the perturbative decay of inflaton.
If we introduce a coupling between inflaton and light matter, which may or may not be the standard model particles,
the inflaton has a finite lifetime and it decays at some epoch in the early universe.
Since we do not know detailed properties of the inflaton, we can arbitrarily choose the inflaton-matter couplings
as long as they do not disturb the inflaton dynamics during inflation.

In this paper we show that inflaton necessarily couples to all scalar fields gravitationally through the cosmic expansion.
We explicitly show that, during the inflaton oscillation regime, 
the Hubble parameter and the cosmic scale factor 
oscillate with (twice) a frequency of inflaton.
That is, the scale factor depends on $\phi^2$, which leads to an effective coupling like $\mathcal L \sim \phi^2 (\partial \chi)^2 /M_P^2$
where $\phi$ is the inflaton, $\chi$ is any scalar field and $M_P$ is the Planck scale.
This leads to ``decay'' or ``annihilation'' of the oscillating inflaton into $\chi$ particles.
A similar phenomenon, called gravitational particle production, was investigated in Ref.~\cite{Ford:1986sy}.
While Ref.~\cite{Ford:1986sy} studied particle production in the transition from the de Sitter space to the radiation or matter dominated universe,
we point out that particle production always occurs in the inflaton oscillation regime.
Although this gravitational interaction is not strong enough to complete the reheating, still it may be a subject of phenomenological interests.
Interestingly, gravitons are also necessarily produced by this process, although its abundance will be below the observable level of
future gravitational wave detectors.

We further discuss the case of non-minimal inflaton coupling models: $f(\phi)R$ gravity.
In such models, the Hubble parameter and the scale factor oscillate more violently with time during the inflaton oscillation regime.
We will explicitly show that the scale factor linearly depends on $\phi$ and hence there arises a gravitationally induced coupling like
$\mathcal L \sim \phi (\partial\chi)^2/M_P$.
This interaction itself can complete the reheating.
 The same result was obtained in Ref.~\cite{Watanabe:2006ku} in a different approach.
 Our method makes it clear that inflaton oscillation explicitly couples to the matter sector inevitably.

 As a by-product, we will show that the inflaton $\phi$ does not always decay into $\chi$ particles
even if it has a coupling like $ \phi (\partial \chi)^2 / M$  in the Lagrangian for $f(\phi)R$ models.
Depending on the choice of $M$, the $\phi$ decay into $\chi$ particles may be forbidden due to some non-trivial cancellation.
Actually, this cancellation necessarily happens in the case of inflaton-graviton coupling.

Before going into the main text, we comment on the terminology of ``gravitational particle production'' 
often used in different contexts and their applicability.
\begin{itemize}
\item Originally Ref.~\cite{Ford:1986sy} called the particle production due to the non-adiabatic change of the background
at the end of inflation as the ``gravitational particle production''.
This is close to what we are going to study, although our main focus is the inflaton-oscillation dominated era after inflation.
We will show that this may be interpreted as the ``decay'' or ``annihilation'' of the inflaton through the gravitational interaction.
In our viewpoint, it is clear that particles lighter than the inflaton are necessarily produced.
\item Quantum fluctuations generated during inflation for nearly massless scalar fields will eventually become
a coherent oscillation mode, which is also called as ``gravitational particle production''~\cite{Felder:1999wt}.
This production mechanism is applicable for scalar fields whose mass lighter than the Hubble parameter during inflation
and we do not consider this production mechanism in this paper.
\end{itemize}

This paper is organized as follows.
In Sec.~\ref{sec:Ein} we study the system of Einstein gravity and minimally-coupled matter
and show the gravitational coupling between inflaton and ordinary matter fields.
In Sec.~\ref{sec:ex} we extend our analysis to the case of non-minimally coupled inflaton.
Sec.~\ref{sec:sum} is devoted to summary and discussion.

\section{Einstein Gravity}  \label{sec:Ein}
\setcounter{equation}{0}

In this section we discuss particle production after inflation in the standard Einstein gravity.
Let us consider the following action
\begin{equation}
	S = \int d^4x \sqrt{-g}\left[ \frac{1}{2}M_P^2R - 
	\frac{1}{2}g^{\mu\nu}\partial_\mu\phi \partial_\nu\phi  -V(\phi)\right] + S_{\rm M},
	\label{S_toy}
\end{equation}
with the Robertson-Walker metric,
\begin{align}
	ds^2 &= -dt^2 + a^2(t) d\vec x^2,
\end{align}
where $g = {\rm det}(g_{\mu \nu})$, $a$ is the scale factor, $R$ is the Ricci scalar, $\phi$ denotes the inflaton, $S_{\rm M}$ collectively represents any other field
and we assume that $S_{\rm M}$ does not explicitly contain the $\phi$ field.
We will see how the inflaton $\phi$ couples to fields in $S_{\rm M}$ 
in the inflaton-oscillation dominated regime in this minimal setup.

\subsection{Background dynamics}

Let us assume that the inflaton dominates the universe.
We consider the homogeneous background dynamics of the inflaton. The Friedmann equation is
\begin{equation}
	3M_{P}^{2}H^2 = \rho_\phi,
\end{equation}
where $H=\dot a/a$ is the Hubble parameter and $\rho_\phi = \dot\phi^2/2 + V$.
The equation of motion of $\phi$ is
\begin{equation}
	\ddot\phi + 3H\dot\phi + V_{,\phi} = 0,
\end{equation}
where $V_{,\phi}=\partial V / \partial\phi$.
From the equation of motion, we obtain
\begin{equation}
	\dot\rho_\phi + 3H \dot\phi^2 = 0.  \label{rhodot}
\end{equation}
Here, we denote $\phi$ as a deviation from the potential minimum.

In the deep oscillation regime $m_\phi \gg H$, where $m_\phi=\sqrt{|V_{,\phi}/\phi|}$ is the inflaton mass scale,\footnote{
Note that this condition corresponds to $\phi/M_{P} \ll 1$.
}
$\rho_{\phi}$ is a good conserved quantity in an inflaton oscillation time scale.
Therefore, from the Virial theorem, we obtain
\begin{equation}
	\left< \dot\phi^2 \right> = n\left< V\right>,
\end{equation}
where we have assumed power-law potential $V \propto \phi^n$ and $\left<\cdots\right>$ means time average
in a period longer than oscillation period but much shorter than the Hubble time scale.
Then we obtain
\begin{equation}
	\left<\dot\rho_\phi \right> + \frac{6n}{n+2}\langle H\rangle \left< \rho_\phi \right> = 0.
\end{equation}
This shows that $\left< \rho_\phi \right> \propto \langle a(t)\rangle^{-6n/(n+2)}$ and $\langle H\rangle = (n+2)/(3nt)$.

This oscillation-averaged picture is often useful when discussing cosmological evolution of the oscillating inflaton field.
However, it does not contain small oscillations in $\rho_\phi$ and $H$, 
which come from the effects neglected in the Virial theorem.  
Although the oscillation amplitude is small, it induces particle production and may have observational impacts in some cases.\footnote{
Such kind of particle production is also briefly discussed in ref.~\cite{Bassett:1997az}.}

To analyze the oscillatory behavior, quantities are divided into oscillation-averaged parts and oscillation parts:
$\rho_\phi = \left<\rho_{\phi}\right> + \delta\rho_\phi$ and $H = \left< H \right> + \delta H$.
The oscillation part satisfies
\begin{equation}
	\delta \dot{H} = \frac{3n}{n+2}\left<H\right>^2 - \frac{\dot\phi^2}{2M_P^2}.
\end{equation}
By noting 
$\dot\phi^2/2+V = \left<\rho_\phi\right> + \delta\rho_{\phi} \simeq 3M_P^2\left<H \right>^2(1 + 2\delta H/\langle H\rangle)$, 
this is rewritten as
\begin{equation}
	\delta \dot{H} + \frac{6n}{n+2}\langle H\rangle \delta H 
	\simeq -\frac{1}{n+2}\frac{1}{M_P^2} \left[\frac{d}{dt} + 3H\right]\left(\phi\dot\phi\right).
	\label{eq:deltaHdot}
\end{equation}
The first term on the l.h.s. is $\delta \dot{H} \sim m_{\phi}\delta H \gg \langle H\rangle \delta H$,
and hence it dominates the second term.
For the same reason, on the r.h.s., the first term gives the dominant contribution.
Then, by integrating this equation, we obtain
\begin{equation}
	\delta H \simeq -\frac{1}{n+2}\frac{\phi\dot\phi}{M_P^2}.
	\label{dH_ana}
\end{equation}
The scale factor $a(t)$ is obtained by integrating $\dot a/a = H = \left< H \right> + \delta H$:
\begin{equation}
	a(t) \simeq \left< a(t)\right>\left( 1-\frac{1}{2(n+2)}\frac{\phi^2-\langle\phi^2\rangle}{M_P^2} \right),~~~
	\left< a(t)\right> \simeq a_i \left( \frac{t}{t_i} \right)^{\frac{n+2}{3n}},
\end{equation}
where quantities with subscript $i$ are evaluated at an arbitrary initial time in the inflaton-oscillation dominated era.\footnote{
Here we neglect terms suppressed by $\sim \langle\phi^2\rangle/M_{P}^2$ 
in the oscillation-averaged scale factor $\langle a(t)\rangle$.
}
This explicitly shows that the Hubble parameter $H$ as well as the scale factor $a(t)$ depend on $\phi^2$ and  oscillate with time.
Note that the relative oscillation amplitude, $\delta H/H \propto \phi/M_{P}$, becomes smaller and smaller as 
the universe expands since the inflaton oscillation amplitude decreases with time.
Thus the oscillation in the scale factor is most efficient just after inflation.

\begin{figure}
\begin{center}
\includegraphics[scale=1.2]{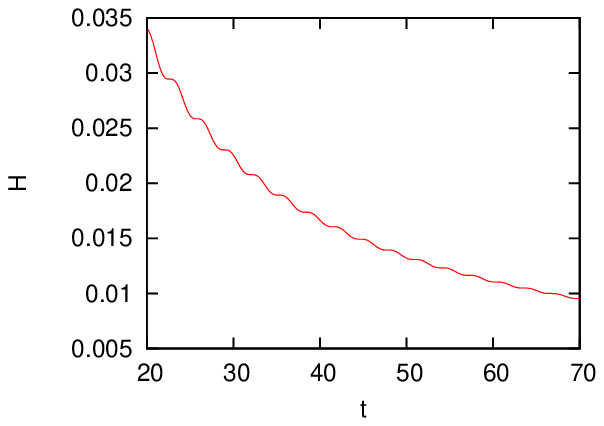}
\includegraphics[scale=1.2]{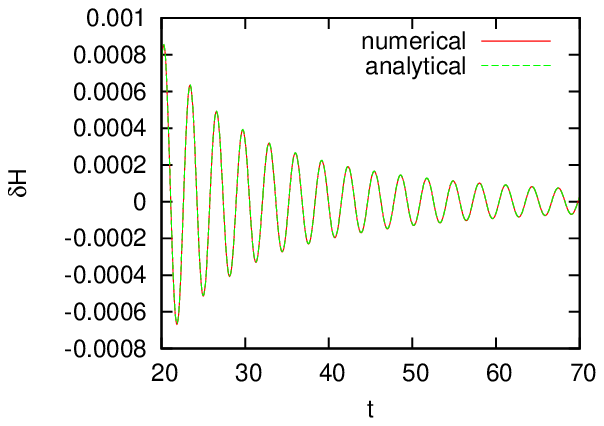}
\caption { 
Time evolution of $H$ (left) and $ \delta H$ (right).
We have compared numerical results and the approximate analytic formula (\ref{dH_ana}) for $\delta H$.
We have taken $m_\phi=1$ in Planck unit.
}
\label{fig:1}
\end{center}
\end{figure}

We have performed numerical calculation to check the above formula.
Fig.~\ref{fig:1} shows time evolution of $H$ (left) and $ \delta H$ (right) for $n=2$.
We have compared numerical results and the approximate analytic formula (\ref{dH_ana}) for $\delta H$.
We have taken $\phi_i=0.1$ and $m_\phi=1$ in Planck unit.

\subsection{Gravitational annihilation of inflaton}

\subsubsection{Scalar}

Now let us discuss the gravitational inflaton decay.
To be concrete, we consider the matter Lagrangian for a real scalar $\chi$:
\begin{equation}
	S_{\rm M} = \int d^4x\sqrt{-g}\left[-\frac{1}{2}g^{\mu\nu}\partial_\mu\chi \partial_\nu\chi -\frac{1}{2}m_\chi^2\chi^2\right],
\end{equation}
where we assume $m_\chi \ll m_\phi$.
As shown above, the scale factor and hence $\sqrt{-g}$ contains $\phi^2$ dependence.
Therefore, neglecting terms including $m_\chi$, the action can be expanded as
\begin{equation}
	S_{\rm M} = \int d\tau d^3x\, \left< a(t)\right>^2\left( 1- \frac{1}{n+2}\frac{\phi^2}{M_P^2} \right) 
	\frac{1}{2}\left[ \chi'^2 - (\partial_{i}\chi)^2 \right],
\end{equation}
where we have used the conformal time $d\tau = dt/a(t)$ and the prime denotes derivative with respect to $\tau$.
This explicitly shows that the inflaton $\phi$ couples to $(\partial \chi)^2$ and $\phi$ (partially) ``decays'' 
or ``annihilates'' into $\chi$ particles.
According to the analysis of particle production under the oscillating background $\phi$~\cite{Shtanov:1994ce,Kofman:1997yn}, it
might be interpreted as the annihilation of the inflaton into $\chi$ particles.
Thus we call this ``gravitational annihilation'' for convenience in the following.

To estimate the production rate, we write down the equation of motion for $\tilde\chi_k \equiv a\chi_k$
where $\chi_k$ denotes the Fourier mode of $\chi$ with comoving wavenumber $k$,
\begin{equation}
	\tilde\chi''_k +\left(k^2 - \frac{a''}{a}\right) \tilde\chi_k = 0 ~~~\rightarrow~~~
	\tilde\chi''_k + \left(k^2-2\left<\mathcal H\right>^2+\frac{\phi'^2}{2M_P^2} \right)\tilde\chi_k= 0,  \label{eqofm_s}
\end{equation}
where $\mathcal H \equiv a'/a=aH$. Therefore, the effective mass of $\tilde\chi$ oscillates rapidly.\footnote{
	If $\chi$ couples to the Ricci scalar as $\mathcal L = \frac{1}{2}f(\chi)R$, the equation of motion becomes
	$\tilde\chi'' - \nabla^2\tilde\chi+(a''/a)(\tilde\chi + 3f_{,\chi} a)=0$.
	Thus if $f(\chi)=-\chi^2/6$, $\tilde\chi$ does not couple to the scale factor or the inflaton in the limit $m_\chi = 0$. 
	This corresponds to the conformal coupling.
}
Particle creation with $k\simeq m_\phi$ occurs and the creation rate was studied e.g. in Ref.~\cite{Shtanov:1994ce}:
\begin{equation}
	\Gamma(\phi\phi \to \chi\chi) \simeq \frac{\mathcal C}{32\pi}\frac{\Phi^2}{M_P^2}\frac{m_\phi^3}{M_P^2},
	\label{Gamma_phi}
\end{equation}
for $n=2$ where $\mathcal C$ is a numerical constant of $\mathcal O(1)$, with $\Phi$ being the inflaton oscillation amplitude.
Since $\Phi$ is decreasing, it never exceeds the Hubble parameter and the reheating is not completed by this gravitational annihilation.\footnote{
It is pointed out that, if reheating ends only through the gravitational particle production in the context of ref.~\cite{Ford:1986sy}, the graviton overproduction can be problematic~\cite{Peebles:1998qn}.}
Still, however, this small annihilation process may yield significant amount of particles.\footnote{
	Note that $\Phi$ in Eq.~(\ref{Gamma_phi}) should be regarded as amplitude measured from the potential minimum of the
	inflaton.
}

Here we comment on the possibility of parametric resonance.
The width of the first resonance band for $\tilde\chi_k$ at $k\simeq m_\phi$ is $\Delta k \sim qm_\phi$ with $q \equiv \Phi^2/2M_P^2$.\footnote{
The width of the instability band may be interpreted as follows.
If one regards the particle production process as an annihilation $\phi\phi\to\chi\chi$,
the dispersion relation of $\chi$ is given by $k_\chi^2=m_\phi^2(1-\phi^2/2M_P^2)$.
Since $\phi$ is oscillating, the momentum of $\chi$ has a finite width of $\Delta k=m_\phi\Phi^2/2M_P^2 = qm_\phi$. 
}
The Hubble expansion removes the produced $\chi$ particles from the instability band in the phase space
within the time interval $\sim q/H$. 
Since the time scale for the exponential growth of the $\chi$ particle is $ \sim (qm_\phi)^{-1}$,
we need  $q^2m_\phi \gg H$ in order for the parametric resonance to occur~\cite{Shtanov:1994ce,Kofman:1997yn}.
This condition is rewritten as $(\Phi/M_P)^3 \gg 1$, which cannot be satisfied as long as $\Phi\ll M_P$.
Thus we neglect the effect of parametric resonance and simply use the production rate (\ref{Gamma_phi}).

Now let us estimate the abundance of $\chi$ particles produced by this gravitational annihilation.
Because of the $\Phi$ dependence, the abundance is dominated by those created just after inflation:
\begin{equation}
	n_\chi (H_{\rm inf}) \simeq \frac{\rho_\phi}{m_\phi}\frac{\Gamma(\phi\phi\to\chi\chi)}{H_{\rm inf}} \simeq \frac{9\mathcal C}{16\pi}H_{\rm inf}^3,
\end{equation}
where $H_{\rm inf}$ is the Hubble parameter just after inflation.
Note that this is of the same order as the number density found in Ref.~\cite{Ford:1986sy} where the particle creation
at the transition from the de Sitter universe to the matter or radiation dominated universe was studied.
Our approach shows that particle creation always occurs in the inflaton-oscillation dominated era and 
makes it clear and explicit that the gravitational particle production is induced by the inflaton-scalar coupling in the Lagrangian.

If $\chi$ is massive (but much lighter than the inflaton) and stable, and assuming that its interaction is so weak that they cannot be thermalized, 
its energy density per entropy density is given by
\begin{equation}
	\frac{\rho_\chi}{s}\simeq \frac{9\mathcal C}{64\pi}\frac{m_\chi T_{\rm R}H_{\rm inf}}{M_P^2}	
	\simeq 8\times 10^{-9}\,{\rm GeV}\, \mathcal C \left( \frac{m_\chi}{10^6\,{\rm GeV}} \right)
	\left( \frac{T_{\rm R}}{10^{10}\,{\rm GeV}} \right)\left( \frac{H_{\rm inf}}{10^{14}\,{\rm GeV}} \right),
	\label{rhochi}
\end{equation}
where $T_{\rm R}$ denotes the reheating temperature, which is completed by some other inflaton interaction terms.
If $\chi$ is absolutely stable, this can be one of the production mechanisms of present dark matter.
On the other hand, if $\chi$ is a moduli which decays into standard model particles through Planck-suppressed interactions,
this is another source of the cosmological moduli problem.\footnote{
	In the context of gravitational particle creation of Ref.~\cite{Ford:1986sy}, see also Refs.~\cite{Chung:1998zb,Felder:1999wt}.
	Ref.~\cite{Nakayama:2011wqa} studied the moduli abundance in the form of coherent oscillation
	induced by the non-adiabatic change of the Hubble parameter.
}

For nearly massless $\chi$ particles such as axion-like particles, the gravitationally produced $\chi$ particles may contribute to dark radiation.
The energy density of such massless $\chi$ particles, normalized by the radiation energy density, is estimated as
\begin{equation}
	\left(\frac{\rho_\chi}{\rho_{\rm rad}}\right)_{T=T_{\rm R}} \simeq \frac{3\mathcal C}{16\pi}\frac{m_\phi H_{\rm inf}^{1/3}H_{\rm R}^{2/3}}{M_P^2}
	\simeq 3\times 10^{-19} \mathcal C \left( \frac{m_\phi}{10^{13}\,{\rm GeV}} \right)
	\left( \frac{T_{\rm R}}{10^{10}\,{\rm GeV}} \right)^{4/3}\left( \frac{H_{\rm inf}}{10^{14}\,{\rm GeV}} \right)^{1/3},
	\label{rhoDR}
\end{equation}
where $H_{\rm R}$ is the Hubble parameter at the completion of the reheating.
Thus its contribution to the present dark radiation energy density is negligibly small.

One comment is in order.
A scalar particle often obtains a Hubble-induced mass through the coupling with Ricci scalar $R$ or the 
Planck-suppressed interactions with inflaton field.
If such a mass term is absent, i.e. if $\chi$ is nearly massless during inflation,
it necessarily develops quantum fluctuations of overhorizon modes 
which in turn will become a coherent oscillation after inflation,
and this may be a dominant contribution to the $\chi$ abundance~\cite{Felder:1999wt}.
On the other hand, if $\chi$ obtains a Hubble mass during inflation, its quantum fluctuation is suppressed.
However, even in such a case, our production mechanism always works as long as $m_\phi \gg H_{\rm inf}$,
which in fact many inflation models including new inflation and hybrid inflation satisfy, 
and this may be a dominant contribution to the $\chi$ abundance.
In this sense, the estimate (\ref{rhochi}) gives a robust lower bound on the $\chi$ abundance
unless $\chi$ is conformally coupled to the Ricci scalar.

\subsubsection{Vector boson and fermion}  \label{sec:fermion}

Next let us consider gravitational particle production of massless vector bosons and fermions. In order to see this, 
we first consider the following action for a massless vector boson:
\begin{align}
S_{\rm M} = -\frac{1}{4}\int d^{4}x \sqrt{-g}g^{\mu\alpha}g^{\nu\beta}F_{\mu\nu}F_{\alpha\beta},
\end{align}
where $F_{\mu\nu}$ is the field strength of the vector boson. If we use the conformal time $d\tau = dt/a(t)$, 
the action can be rewritten as
\begin{align}
S_{\rm M} = -\frac{1}{4}\int d\tau d^{3}x \eta^{\mu\alpha}\eta^{\nu\beta}F_{\mu\nu}F_{\alpha\beta},
\end{align}
where $\eta_{\mu\nu}$ is the metric in the Minkowski space-time and the summations are taken over 
$\{\tau, x_{1}, x_{2}, x_{3}\}$.
From this, it is obvious that the massless vector boson does not couple to the scale factor.
Note that this discussion holds also for non-abelian gauge fields.
The same thing happens for the case of a massless fermion. To see this we consider the following action:
\begin{align}
S_{\rm M} = -\int d^4x e\bar{\psi}e^{\mu}_{a}\gamma^{a}\mathcal{D}_{\mu}\psi,
\end{align}
where $\psi$ is the spinor field, $e^{\mu}_{a}$ is the vierbein and $e \equiv {\rm{det}}(e_{\mu}^{a}) = \sqrt{-g}$.
Here we use the Greek letters $\mu, \nu, ...$ as the Einstein indices and the Latin letters $a, b, ...$ as 
the local Lorentz indices.
Moreover, $\mathcal{D}_{\mu}$ is the covariant derivative which is defined as
\begin{align}
\mathcal{D}_{\mu} = \partial_{\mu} + \frac{1}{4}\omega_{\mu}^{\ ab}\gamma_{[a}\gamma_{b]},
\end{align}
where the spin connection $\omega_{\mu}^{\ ab}$ is given as
\begin{align}
\omega^{\ ab}_{\mu} = 2e^{\nu [a}\partial_{[\mu}e_{\nu]}^{\ b]} - e^{\nu[a}e^{b]\sigma}e_{\mu c}\partial_{\nu}e_{\sigma}^{\ c}.
\end{align}
By using the conformal time and rescaling the spinor field as $\tilde{\psi} \equiv a^{3/2}(t)\psi$,  the action can be 
rewritten as
\begin{align}
S_{\rm M} = -\int d\tau d^{3}x \bar{\tilde{\psi}}\delta^{\mu}_{a}\gamma^{a}\partial_{\mu}\tilde{\psi},
\end{align}
where the summation for $\mu$ is taken over $\{\tau, x_{1}, x_{2}, x_{3}\}$.
Therefore, the scale factor does not couple to the massless fermion.
Note that these facts are closely related to the Weyl invariance of massless vector bosons and fermions.
In other words, this annihilation process is applied to any fields lighter than the inflaton if
they are not Weyl invariant.\footnote{
	See Refs.~\cite{Giudice:1999yt} for the gravitino production.
}
For example, if vector bosons and fermions are massive, they are coupled to the scale factor 
and hence produced by the inflaton oscillation.

\subsubsection{Graviton}

Finally, let us see the inflaton-graviton coupling.
The graviton is defined as the transverse-traceless part of the metric perturbation:
\begin{equation}
	ds^2 = -dt^2 + a(t)^2(\delta_{ij} + h_{ij})dx^idx^j,
	\label{eq:graviton}
\end{equation}
where $h_{ij}$ satisfies $\partial_i h_{ij}=h_{ii}=0$.
Expressing the two helicity modes by $h_{\lambda}$\,$(\lambda=+,\times)$, it is known that
the equation of motion of the graviton is the same as that of the minimally-coupled massless scalar field.
Thus $\tilde h_\lambda\equiv ah_\lambda$ satisfies the same equation as (\ref{eqofm_s}).
It means that the graviton production rate is given by
\begin{equation}
	\Gamma(\phi\phi \to hh) \simeq \frac{\mathcal C}{16\pi}\frac{\Phi^2}{M_P^2}\frac{m_\phi^3}{M_P^2}.
\end{equation}
The abundance of graviton, or the gravitational wave, is given by Eq.~(\ref{rhoDR}).
The present frequency of the gravitational wave is
\begin{equation}
	f_{\rm GW} \simeq \frac{m_\phi a(H_{\rm inf})}{2\pi} \simeq 2\times 10^5\,{\rm Hz} 
	 \left( \frac{m_\phi}{10^{13}\,{\rm GeV}} \right)
	\left( \frac{T_{\rm R}}{10^{10}\,{\rm GeV}} \right)^{1/3}\left( \frac{H_{\rm inf}}{10^{14}\,{\rm GeV}} \right)^{-2/3},
\end{equation}
where we have taken the present scale factor to be equal to one.
The present gravitational wave spectrum in terms of $\Omega_{\rm GW}$ 
has a peak at this frequency and scales as $f^{-1/2}$ for higher frequencies.
Around this frequency range, the abundance of gravitational wave is typically below the observable level of future space laser interferometers~\cite{Seto:2001qf}.

\section{Extended Gravity} \label{sec:ex}
\setcounter{equation}{0}

In the previous section we have shown that, in the Einstein gravity, 
the inflaton necessarily couples to scalar fields and graviton through the $\phi$ dependence of the scale factor,
which is proportional to $\phi^{2}$.
In this section, we discuss particle production in the extended gravity,
since it sometimes happens that the Hubble parameter as well as the scale factor oscillates more violently~\cite{Arbuzova:2011fu,Jinno:2013fka}. This means that the scale factor linearly depends on $\phi$. 
Therefore, 
we will expand $H$ and $a(t)$ at the linear order in $\phi$ and neglect $\phi^2$ dependence in this section.

Let us consider the following action:
\begin{equation}
	S = \int d^4x \sqrt{-g}\left[ \frac{1}{2}f(\phi)R
	- \frac{1}{2}g^{\mu\nu}\partial_\mu\phi \partial_\nu\phi  -V(\phi) \right] + S_{\rm M},
	\label{S_toy}
\end{equation}
where $f(\phi)$ is an arbitrary function of $\phi$, which can be expanded as
\begin{equation}
	f(\phi) = M_{P}^2\left(1 + c_1\frac{\phi}{M_{P}} + \cdots\right).
\end{equation}
Here we take $f(0) = M_{P}^2$
without loss of generality by redefining $\phi$ as a deviation from the potential minimum.

\subsection{Background dynamics}  \label{sec:homo}

Let us first consider the homogeneous mode of $\phi$ which undergoes a coherent oscillation around the potential minimum:
$\phi = \phi(t)$. 
Let us compute $\phi$-dependence of the scale factor $a(t)$.
From the Einstein equation, we obtain
\begin{equation}
	3fH^2 = \rho_\phi - 3H\dot f,  \label{F_toy}
\end{equation}
where $\rho_\phi \equiv \dot\phi^2/2 + V$ and
\begin{equation}
	(3H^2 + 2\dot H) f + \ddot f + 2H\dot f = -\frac{1}{2}\dot\phi^2 + V.  \label{Fried2}
\end{equation}
The equation of motion of $\phi$ reads
\begin{equation}
	\ddot \phi + 3H\dot \phi+ V_{,\phi} - 3f_{,\phi}(2H^2 + \dot H)=0.  \label{phi_toy}
\end{equation}
One of these three equations is actually redundant.

It should be noticed that the Hubble parameter $H$ is an oscillating function with time,
as opposed to the case of the Einstein gravity.
To see this, we expand $H$ and $\phi$ as $H = H_0 + H_1$ and $\phi = \phi_0 + \phi_1$,
where the subscript $0$ represents the solution in the Einstein gravity limit $f(\phi) = M_{P}^2$, and we regard the 
effect of non-minimal coupling $c_1$ as a small perturbation.\footnote{
Note that this expansion is different from what we have done in the previous section.}
Thus we have $H_0 = (n+2)/(3nt)$ for $V \propto \phi^n$.
We also consider the deep oscillation regime: i.e., $H \ll m_\phi$, with $m_\phi$ being the mass scale of $\phi$.

From Eq.~(\ref{F_toy}), we obtain
\begin{align}
	H_0^2 &= \frac{\rho_{\phi0}}{3M_{P}^{2}}, \\
	H_{1} &= \frac{\rho_{\phi1}}{6M_{P}^2H_{0}} - \frac{c_{1}}{2M_{P}}\left(\dot{\phi}_{0} + H_{0}\phi_{0}\right),
	\label{eq:3c1H02}
\end{align}
where $\rho_{\phi1}=\dot\phi_0\dot\phi_1 + V_{,\phi 0}\phi_1$.
In order to estimate the magnitude of $\rho_{\phi1}$, we rewrite (\ref{phi_toy}) as
\begin{equation}
	\dot\rho_{\phi1}+ 6H_0 \dot\phi_0\dot\phi_1+3H_1\dot\phi_0^2 -3c_1M_{P}\dot\phi_0(2H_0^2 + \dot H)=0.
\end{equation}
This is further rewritten as
\begin{equation}
	\dot\rho_{\phi1}+ 6H_0 \dot\phi_0\dot\phi_1
	+\frac{\dot\phi_0^2}{2M_{P}^{2}H_0}\rho_{\phi1}
	=\frac{2c_1}{M_{P}}\rho_{\phi0}\dot\phi_0.
\end{equation}
The second and third terms on the l.h.s. are of the order of $\sim H \rho_{\phi1}$, and 
are subdominant compared to the first term.
Thus integrating this equation, we obtain $\rho_{\phi1}\sim c_{1}M_{P}H_{0}^{2}\phi_{0}$,
since $\dot\phi$ is an oscillating function 
with time scale $\sim 1/m$.
Therefore, on the r.h.s. of Eq.~\eqref{eq:3c1H02}, the second term dominates over the other terms.
As a result, we obtain
\begin{equation}
	H_1 \simeq -\frac{c_1}{2M_{P}}\dot\phi_0.
	\label{H_ana}
\end{equation}
Integrating $\dot a/a = H$, we find the scale factor 
\begin{equation}
	a(t) \simeq a_i\left( \frac{t}{t_i} \right)^{\frac{n+2}{3n}}\left(1- \frac{c_1}{2}\frac{\phi}{M_{P}} \right) 
	\equiv  a_0(t) \left(1- \frac{c_1}{2}\frac{\phi}{M_{P}} \right).
	\label{a_ana}
\end{equation}
These results show that the Hubble parameter $H$ as well as the scale factor $a$ explicitly depend on $\phi(t)$ and hence rapidly oscillating functions with time.
The combination $a^2(t) f(\phi)$ becomes
\begin{equation}
	a^2(t) f(\phi) \simeq M_{P}^{2}a_0(t)^2.
\end{equation}
Thus the $\phi$ dependence cancels out in the combination $a^2(t) f(\phi)$.

\begin{figure}
\begin{center}
\includegraphics[scale=1.2]{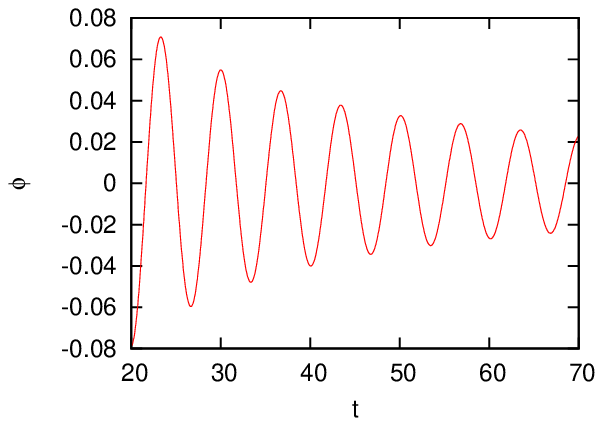}
\includegraphics[scale=1.2]{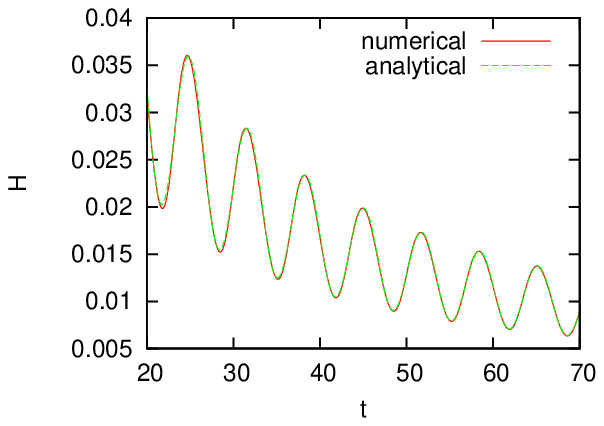}
\includegraphics[scale=1.2]{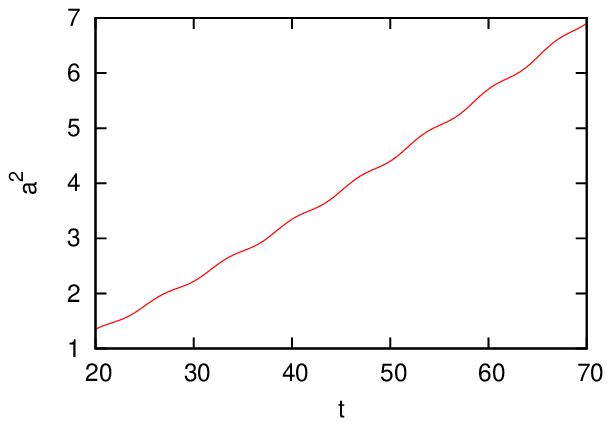}
\includegraphics[scale=1.2]{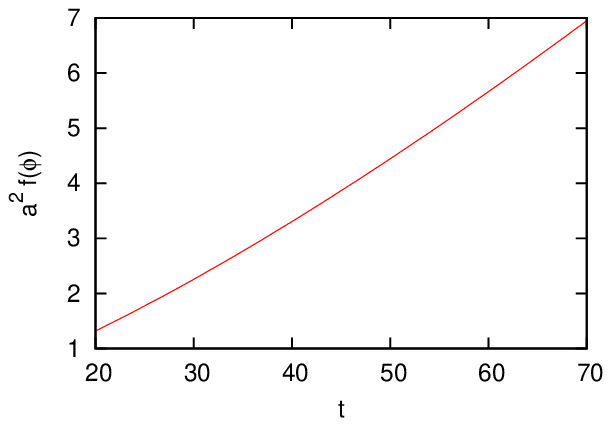}
\caption { 
Time evolution of $\phi(t)$ (top left), $H$ (top right), $a^2$ (bottom left) and $a^2 f(\phi)$ (bottom right).
We have compared numerical results and approximate analytic formula (\ref{H_ana}) for $H$.
We have taken $c_1=0.3$ and $m_\phi=1$ in Planck unit.
}
\label{fig:2}
\end{center}
\end{figure}

We have performed numerical calculation to confirm the above considerations.
Fig.~\ref{fig:2} shows time evolution of $\phi(t)$ (top left), $H$ (top right), $a^2$ (bottom left) and $a^2 f(\phi)$ (bottom right)
for $n=2$.
We have taken $\phi_i=0.1$, $c_1=0.3$ and $m_\phi=1$ in Planck unit.
We have compared numerical results and approximate analytic formula (\ref{H_ana}) for $H$ and (\ref{a_ana}) for $a^2$.
It is clearly seen that $H$ as well as the scale factor $a$ oscillates with time,
but the combination $a^2 f(\phi)$ does not.

\subsection{Gravitational decay of inflaton}

\subsubsection{Scalar}
\label{subsec_scalar}

First let us consider the inflaton decay into a real scalar $\chi$. We introduce the following term in the action
\begin{equation}
	S_{\rm M}= \int d^4x \sqrt{-g}\left[ - \frac{1}{2}h(\phi)g^{\mu\nu}\partial_\mu\chi \partial_\nu\chi \right],
	\label{nonmin_scalar}
\end{equation}
where $h(\phi)$ is an arbitrary function of $\phi$, which is expanded as $h(\phi)= 1 + d_1\phi/M_{P} + \cdots$.
The reason for the introduction of this non-minimal coupling will become clear soon.
Naively, this term leads to the decay rate of the inflaton as $\Gamma(\phi\to \chi\chi) = d_1^2m_\phi^3/(128\pi M_{P}^{2})$.
But this naive expectation is not correct as we will see below.

Clearly $h(\phi)$ is an oscillating function and it might seem that it induces particle creation of $\chi$.
But we must be also careful on the prefactor $\sqrt{-g}$, which could cancel the oscillation of $h(\phi)$.
More precisely, by using the conformal time and metric matter action becomes
\begin{equation}
	S_{\rm M} = \int d\tau d^3x \,a^2(t) h(\phi)\frac{1}{2}\left[\chi'^2 - (\partial_{i} \chi)^2\right].  \label{S_toy2}
\end{equation}
Thus it is the combination $a^2(t) h(\phi)$ that determines the particle creation rate of $\chi$.
By using the result obtained in the previous section, the combination $a^2(t) h(\phi)$ becomes
\begin{equation}
	a^2(t) h(\phi) \simeq a_0(t)^2\left[1+ \left(d_1- c_{1}\right)\frac{\phi}{M_{P}} \right].
\end{equation}
Thus the $\phi$ dependence vanishes for $d_1 = c_1$, i.e., $f(\phi) = M_{P}^{2} h(\phi)$.
In this case, $\phi$ does not couple to $\chi$, hence no $\chi$ particle production is induced by the $\phi$ oscillation
except for the effect studied in the previous section.
In the present toy model, there is no reason to expect $f(\phi) = M_{P}^{2} h(\phi)$ and hence the decay occurs unless there is a tuning,
but later we will see that this tuning automatically happens if we regard $\chi$ as the graviton.

Having revealed the $\phi$-$\chi$ coupling, we can calculate the decay rate of $\phi$ as
\begin{equation}
	\Gamma(\phi\to \chi\chi) = \frac{m_\phi^3}{128\pi M_{P}^{2}}\left(d_1- c_{1}\right)^2.
\end{equation}
It coincides with the result in Ref.~\cite{Watanabe:2006ku} for $d_1=0$.
It is noticed that even for $d_1=0$, i.e., no explicit coupling between $\phi$ and $\chi$,
the rapid oscillation of the scale factor leads to the efficient decay of the inflaton.

\subsubsection{Vector boson and fermion}

Next let us consider vector bosons and fermions:
\begin{align}
	S_{\rm M} = -\int d^{4}x \sqrt{-g}\left[h_G(\phi) \frac{1}{4}g^{\mu\alpha}g^{\nu\beta}F_{\mu\nu}F_{\alpha\beta} 
	+ h_F(\phi)\bar{\psi}e^{\mu}_{a}\gamma^{a}\mathcal{D}_{\mu}\psi \right],
\end{align}
where $h_G(\phi)$ and $h_F(\phi)$ are arbitrary functions of $\phi$,
which can be expanded as $h_G(\phi) = 1+ d_G \phi/M_{P} + \cdots$ and $h_F(\phi)= 1+ d_F \phi/M_{P} + \cdots$.
As shown in Sec.~\ref{sec:fermion}, by using the conformal time and appropriate rescaling of the fields,
they can be rewritten as the form that does not include the scale factor:
\begin{equation}
	S_{\rm M} = -\int d\tau d^{3}x \left[ h_G(\phi) \frac{1}{4} \eta^{\mu\alpha}\eta^{\nu\beta}F_{\mu\nu}F_{\alpha\beta}
		+ h_F(\phi) \bar{\tilde{\psi}}\delta^{\mu}_{a}\gamma^{a}\partial_{\mu}\tilde{\psi} \right].
\end{equation}
Thus the oscillation in the scale factor does not affect the decay rate of $\phi$ into massless vector bosons and fermions.
The decay rate into vector bosons is given by
\begin{equation}
	\Gamma(\phi \to gg) = \frac{N}{64\pi M_{P}^2} d_G^2 m_\phi^3,
\end{equation}
where $N$ denotes the number of vector bosons.
The decay rate into massless fermions vanishes.

\subsubsection{Graviton}  \label{sec:graviton}

Here we apply our results to the inflaton-graviton coupling.
Assuming that the background is homogeneous and expanding only the tensor part as Eq.~\eqref{eq:graviton},
we obtain the graviton action as\footnote{
	Terms proportional to $(h_{ij})^2$ is cancelled out by using the background equation (\ref{Fried2}).
}
\begin{equation}
	S 
	= \int d\tau d^3x \,a(t)^2  f(\phi)\frac{1}{8}\left[ h_{ij}'^2- (\partial_{l} h_{ij})^2 \right],
\end{equation}
where $h_{ij}$ satisfies $h_{ii}= \partial_i h_{ij}=0$.
As shown in Sec.~\ref{sec:homo}, the combination $a(t)^2 f(\phi)$ does not depend on $\phi$.
The $\phi$-dependence of scale factor $a(t)$ through the Friedmann equation cancels the $\phi$-dependence of $f(\phi)$.
Hence there does not appear an inflaton-graviton coupling like $\phi(\partial h)^2$ in $f(\phi) R$ models.

\subsection{Coupling of the fluctuation}  \label{sec:fluc}

In Sec.~\ref{subsec_scalar} 
we have seen that the homogeneous coherently oscillating scalar couples to scalar field $\chi$ unless $f(\phi)=M_P^2h(\phi)$.
In this section we consider the same matter action as that in Sec.~\ref{subsec_scalar}.
Let us assume that $\phi$ sits at the potential minimum $\bar{\phi}$ and consider the fluctuation $\delta\phi(t,\vec x)$ around it.
Hence the background part $\bar\phi$ satisfies the following equations:
\begin{align}
	3H^{2}f(\bar{\phi}) &= V(\bar{\phi}) , \\
	V_{,\phi}(\bar{\phi}) - 6H^{2}f_{,\phi}(\bar{\phi}) &=0.
	\label{eq:bgeom2}
\end{align}
If $V(\bar\phi)=0$, we are dealing with Minkowski space, 
while $V(\bar\phi)>0$ corresponds to de Sitter space.
We want to know the coefficient of coupling like $\sim \delta\phi (\partial\chi)^2$ in the Lagrangian.
This is a non-trivial task because of the mixing between $\delta\phi$ and the fluctuation of scalar part of the metric
due to the non-minimal coupling.

We make use of the ADM formalism \cite{Arnowitt:1962hi} 
to deal with the metric fluctuation and specify the physical degrees of freedom:\footnote{
The analysis of fluctuations during inflationary regime using the ADM formalism is done e.g. by Ref.~\cite{Maldacena:2002vr}.
}
\begin{align}
	ds^{2} = -N^{2}dt^{2} + \gamma_{ij}\left(dx^{i} + \beta^{i}dt\right)\left(dx^{j} + \beta^{j}dt\right),
	\label{ADM}
\end{align}
where $N$ is the lapse function, $\beta^{i}$ is the shift vector and $\gamma_{ij}$ is the metric of space.
The inverse metric is given as 
\begin{align}
	g^{\mu\nu} = \frac{1}{N^{2}}\begin{pmatrix}- 1 &\beta^{i} \\
	\beta^{i} &N^{2}\gamma^{ij} - \beta^{i}\beta^{j} \\
\end{pmatrix}.
\end{align}
In this parametrization, the Ricci scalar can be decomposed as
\begin{align}
	R = R^{(3)} + \frac{1}{N^2}\left(E^{ij}E_{ij} - E^{2}\right) + 2 \nabla_{\mu}\left(\frac{E}{N}n^{\mu}\right) - \frac{2}{N}\Delta^{(3)}N,
\label{eq:Ricci}
\end{align}
where the superscript $(3)$ denotes the covariant quantity with respect to the spacial metric $\gamma_{ij}$.
In addition, $E_{ij}$ is related to the extrinsic curvature $K_{ij}$ as
\begin{align}
	E_{ij} = N K_{ij} = \frac{1}{2}\left(\dot{\gamma}_{ij} - \nabla^{(3)}_{i}\beta_{j} - \nabla^{(3)}_{j}\beta_{i} \right),
\end{align}
and $E = \gamma^{ij}E_{ij}$. Moreover, $n^{\mu}$ is the unit normal vector of the time-like hyper-surface:
\begin{align}
	n^{\mu} = \frac{1}{N}\begin{pmatrix}
	1, &-\beta^{i} \\
\end{pmatrix}.
\end{align}
By using Eq.~\eqref{eq:Ricci}, the Lagrangian can be rewritten as follows:
\begin{align}
	S = \int d^{4}x \sqrt{\gamma}&\left[\frac{1}{2}f(\phi)\left(NR^{(3)}+\frac{1}{N}\left(E^{ij}E_{ij} - E^{2}\right)\right)\right. \nonumber \\ 
	&\left. -f^{\prime}(\phi)\frac{E}{N}\left(\dot{\phi}-\beta^{i}\partial_{i}\phi\right) 
	- N \gamma^{ij}\nabla^{(3)}_{i}\partial_{j}f(\phi) \right. \nonumber \\
	&\left. + \frac{1}{2N}\left(\dot{\phi} - \beta^{i}\partial_{i}\phi\right)^{2} 
	- \frac{N}{2}\gamma^{ij}\partial_{i}\phi\partial_{j}\phi - NV(\phi)\right. \nonumber \\
	&\left. + \frac{h(\phi)}{2N}\left(\dot{\chi} - \beta^{i}\partial_{i}\chi\right)^{2} 
	- \frac{Nh(\phi)}{2}\gamma^{ij}\partial_{i}\chi\partial_{j}\chi \right].
	\label{eq:S-expnd}
\end{align}
The equations of motion for 
the lapse function and shift vector give the constraints among the fields. The constraint equations 
derived from the Lagrangian \eqref{eq:S-expnd} are
\begin{align}
\rm{lapse}:\ 0 =& 
\frac{1}{2}f(\phi)\left(R^{(3)} - \frac{1}{N^{2}}(E^{ij}E_{ij} - E^{2})\right) 
+ \frac{f_{,\phi}E}{N^{2}}(\dot{\phi}-\beta^{i}\partial_{i}\phi)  \nonumber \\
&- \gamma^{ij}\nabla^{(3)}_{i}\partial_{j}f(\phi)
 - \frac{1}{2N^{2}}\left(\dot{\phi} - \beta^{i}\partial_{i}\phi\right)^{2}
 - \frac{1}{2}\gamma^{ij}\partial_{i}\phi\partial_{j}\phi - V(\phi), 
 \label{eq:constraints1} \\
\rm{shift}:\ 0 =& 
\nabla^{(3)}_{i}\left[\frac{1}{N}f(\phi)(E^{i}_{j} - \delta^{i}_{j}E)\right] 
+ \frac{f_{,\phi}E}{N}\partial_{j}\phi  \nonumber \\
&-\nabla^{(3)}_{j}\left[\frac{f_{,\phi}}{N}(\dot{\phi}-\beta^{i}\partial_{i}\phi)\right]
- \frac{1}{N}(\dot{\phi}-\beta^{i}\partial_{i}\phi)\partial_{j}\phi,
\label{eq:constraints2}
\end{align}
where we omit $\chi$-dependent terms by assuming that $\chi$ is a small quantity.\footnote{
This treatment is justified since we solve the constraint equations only to 
first order in perturbations.
}

We now expand the scalar field and metric around the background part as follows:\footnote{
It completely fixes the gauge condition for the perturbations.
}
\begin{align}
	N &= 1+\alpha, \nonumber \\
	\beta^{i} &= \beta^{i}_{\rm{T}}, \ \ \partial_{i}\beta^{i}_{\rm{T}} = 0, \nonumber \\
	\gamma_{ij} &= e^{2\sigma + 2\zeta}(e^{h})_{ij}, \ \ \partial_{i}h_{ij} = h_{ii} = 0, \nonumber \\
	\phi &= \bar{\phi} + \delta\phi, 
	\label{eq:gauge1}
\end{align}
and treat $\alpha$, $\beta_{\rm{T}}^{i}$, $\zeta$, $h$, $\delta\phi$ as perturbations. 
The background scale factor is given by $a(t)=e^{\sigma}$.
It is enough to solve the constraint equations \eqref{eq:constraints1} and \eqref{eq:constraints2}
to first order in the perturbations in order to obtain cubic terms 
in the perturbations. This is because contributions from higher order terms vanish due to the background equations of motion.
Then, for our purpose, it is enough to expand $f(\phi)$ to first order in the perturbations:
\begin{align}
	f(\phi) &\simeq f(\bar{\phi}) + f_{,\phi}(\bar{\phi})\delta\phi = M_{P}^{2}\left(1 + c_1\frac{\delta\phi}{M_{P}}\right),
\end{align}
where we set $f(\bar{\phi}) \equiv M_{P}^{2}$ and $f_{,\phi}(\bar{\phi}) \equiv c_1M_{P}$ in the second line. Only terms up 
to this order is relevant for the constraint equations. 
It is straightforward to solve Eqs.~\eqref{eq:constraints1} and \eqref{eq:constraints2}
by substituting the perturbative expansion \eqref{eq:gauge1}. 
The solutions are
\begin{align}
	\alpha = \zeta &= -\frac{c_1}{2}\frac{\delta\phi}{M_{P}}, \nonumber \\
	\beta_{\rm{T}}^{i} &= 0,
	\label{eq:sol1}
\end{align}
where we used the background equations of motion \eqref{eq:bgeom2} to obtain this result.
This relates the metric fluctuations $\alpha$ and $\zeta$ to $\delta\phi$.

Then turning back to the Lagrangian \eqref{eq:S-expnd}, the last term is expanded up to first order in the perturbations as
\begin{align}
	S &= \int d^4x\, a(t)^3\frac{1}{2}\left[ \left(d_1 \frac{\delta\phi}{M_{P}} + 3\zeta-\alpha\right)\dot\chi^2 
	- \left(d_1\frac{\delta\phi}{M_{P}}+\zeta+\alpha\right)
	\frac{(\partial_{i}\chi)^2}{a^2}  \right] \nonumber \\
	&= \int d^4x\, a(t)^3\frac{1}{2}\left(d_1- c_{1}\right)\frac{\delta\phi}{M_{P}} \left[ \dot\chi^2 - \frac{(\partial_{i}\chi)^2}{a^2} \right].
\end{align}
This agrees with the $\phi$-$\chi$ coupling obtained in the previous subsection in the case of homogenous oscillating 
$\phi$ field. In particular, the coupling vanishes for $d_1=c_1$.
Moreover, we can see that the coupling of the scalar fluctuation with gravitons also vanishes. 
In the ADM parametrization, the graviton action in the form of scalar-graviton-graviton is written as
\begin{equation}
	S = \int d^4x\,a(t)^3 \frac{M_{P}^{2}}{8}\left[ \left(c_{1}\frac{\delta\phi}{M_{P}}+3\zeta-\alpha\right)(\dot h_{ij})^2 
	- \left(c_{1}\frac{\delta\phi}{M_{P}}+\zeta+\alpha\right)\frac{(\partial_{l} h_{ij})^2}{a^2} \right].
\end{equation}
By substituting the solution of the constraint equation (\ref{eq:sol1}), we find that the scalar-graviton-graviton interaction vanishes.
Therefore, the non-minimal coupling like $f(\phi) R$ does not induce the inflaton decay into gravitons.
This result is consistent with that of the previous section.
However, we note that there still remains a coupling $\sim \phi^2 (\partial h)^2$ as studied in Sec.~\ref{sec:Ein}.

\subsection{Conformal transformation}

The calculations so far indicate that the $\phi$ decay into $\chi$ pair is forbidden if $f(\phi) = M_{P}^{2}h(\phi)$.
These results may be understood as a conformal transformation.
We have analyzed the toy model in the Jordan frame, but we can move to the Einstein frame in which the
gravitational action takes the Einstein-Hilbert form by the following conformal transformation:
\begin{align}
	g_{{\rm E} \mu\nu} = \Omega^2 g_{{\rm J} \mu\nu},~~~\Omega^2 = f(\phi_{\rm J})/M_{P}^{2}.
\end{align}
Here the subscript J and E denote the quantities in the Jordan and Einstein frame, respectively.
In this subsection, we keep the subscript J to make the discussion clear.
By the conformal transformation, the Ricci scalar transforms as
\begin{equation}
	R_{\rm J} = \Omega^2\left( R_{\rm E} +6\Box_{\rm E}\ln \Omega-6g_{\rm E}^{\mu\nu}
		\partial_{\mu} \ln\Omega\partial_{\nu} \ln\Omega \right).
\end{equation}
Then the action (\ref{S_toy}) combined with (\ref{nonmin_scalar}) is rewritten as
\begin{equation}
	S = \int d^4x \sqrt{-g_{\rm E}}\left[ \frac{1}{2}M_{P}^{2} R_{\rm E} 
	- \frac{1}{2}g_{{\rm E}}^{\mu\nu}\partial_\mu \phi_{\rm E} \partial_\nu \phi_{\rm E}  -V_{\rm E}(\phi_{\rm E})
	- \frac{1}{2}\frac{M_{P}^2 h(\phi_{\rm J})}{f(\phi_{\rm J})} 
	g_{{\rm E}}^{\mu\nu}\partial_\mu \chi_{\rm J} \partial_\nu \chi_{\rm J}  \right],
\end{equation}
where
\begin{align}
	V_{\rm E}(\phi_{\rm E}) = \frac{V_{\rm J}(\phi_{\rm J})}{\Omega^4}, \\
	d\phi_{\rm E}=M_{P}\sqrt{\frac{2f + 3f_{,\phi}^2}{2f^2} }d\phi_{\rm J}.
\end{align}
Thus it is evident that the $\phi$-$\chi$ interaction vanishes for $f(\phi) = M_{P}^{2}h(\phi)$.

\section{Summary and Discussion}  \label{sec:sum}
\setcounter{equation}{0}

We have shown that the inflaton necessarily couples to any scalar field
as long as it is not conformally coupled to the Ricci scalar through the cosmic expansion 
during the inflaton-oscillation regime.
The point is that the Hubble parameter and the cosmic scale factor have small oscillatory features with (twice) the frequency of the inflaton oscillation.
It induces an effective coupling like $\mathcal L\sim \phi^2(\partial\chi)^2/M_P^2$ that leads to the
``annihilation'' of the inflaton coherent oscillation into $\chi$ particles.
Although the rate of particle creation is rapidly decreasing with time compared with the Hubble parameter and hence
the reheating is not completed by this gravitational annihilation, still it can have phenomenological consequences such as
the heavy dark matter production or the cosmological moduli problem.
Here we stress that any scalar particle lighter than the inflaton is produced by this process,
even if it has a Hubble-induced mass or a mass larger than the Hubble scale during inflation.
The same particle production also applies to the graviton;
high frequency gravitational waves are necessarily produced after inflation.

Moreover, we have investigated the case of inflaton non-minimal coupling: $f(\phi) R$ model.
In this model, the Hubble parameter and the cosmic scale factor oscillate more violently with the frequency of inflaton oscillation.
That induces an effective coupling like $\mathcal L\sim \phi(\partial\chi)^2/M_P$ that leads to the decay of inflaton into $\chi$ particles.
The decay rate can be large enough to complete the reheating.
However, we also found that such a linear coupling to the graviton is automatically cancelled.

Our finding may be understood as particle production in a time-dependent background in a broad sense.
In the context of preheating~\cite{Shtanov:1994ce,Kofman:1997yn}
it is well known that the explicit coupling like $\mathcal L\sim \phi^2\chi^2$ or $\phi\bar\psi\psi$ leads to efficient particle production.
We emphasize that even if there are no explicit couplings among light particles and the inflaton,
they actually couple to each other gravitationally.
More formally, the factor $\sqrt{-g} = a(t)^3$ in the action is directly related to the inflaton dynamics through the Friedmann equation,
which may also be regarded as the constraint equation that connects the metric to the inflaton.
In the Einstein gravity, the oscillatory behavior of the scale factor is rather mild,
but in some extended gravity models, the oscillation might be more drastic
so that the gravity-induced couplings are strong enough to complete the reheating.

Although in this paper we have considered the inflaton, similar discussions are applied to 
any coherently-oscillating scalar field, because the scale factor oscillates 
even if the energy density of the scalar field is subdominant in the universe.

\section*{Acknowledgments}

This work was supported by the Grant-in-Aid for Scientific Research on Scientific Research (A) (No.26247042 [KN]),
Young Scientists (B) (No.26800121 [KN]) and Innovative Areas 2603 ([KN]). 
The work of R.J. and K.M. was supported in part by JSPS Research Fellowships for Young Scientists.
The work of Y.E. and R.J. was also supported in part by the Program for Leading Graduate Schools, MEXT, Japan.

\appendix

\section{Generalization of the Inflaton Sector}
In Sec. \ref{sec:fluc}, we consider the couplings between the fluctuation $\delta\phi(t, \vec{x})$ and $\chi$
for the minimal kinetic term of $\phi$.
The same analysis can be done for a more general form of $\phi$ sector. In this section,
we consider the following Lagrangian:
\begin{align}
S = \int d^{4} \sqrt{-g} \left[\frac{1}{2}f(\phi)R + K(\phi, X) - \frac{h(\phi)}{2}g^{\mu\nu}\partial_{\mu}\chi\partial_{\nu}\chi\right],
\end{align}
where $X  = -g^{\mu\nu}\partial_{\mu}\phi\partial_{\nu}\phi/2$. Note that this Lagrangian includes a variety of extended
gravity models such as $f(R)$ models.

As in Sec. \ref{sec:fluc}, let us assume that $\phi$ sits at the potential minimum and consider the fluctuation $\delta\phi(t,\vec x)$ around it. The background part $\bar\phi$ satisfies the following equations:
\begin{align}
	0 &= K_{,\phi}(\bar{\phi}, 0) + 6H^{2}f_{,\phi}(\bar{\phi}), 
	\label{eq:bgeom-gene1} \\
	0 &= K(\bar{\phi}, 0) + 3H^{2}f(\bar{\phi}).
	\label{eq:bgeom-gene2}
\end{align}
By using the ADM parametrization, the Lagrangian can be rewritten as
\begin{align}
	S = \int d^{4}x \sqrt{\gamma}&\left[\frac{1}{2}f(\phi)\left(NR^{(3)}+\frac{1}{N}\left(E^{ij}E_{ij} - E^{2}\right)\right)\right. \nonumber \\ 
	&\left. -f^{\prime}(\phi)\frac{E}{N}\left(\dot{\phi}-\beta^{i}\partial_{i}\phi\right) 
	- N \gamma^{ij}\nabla^{(3)}_{i}\partial_{j}f(\phi) + NK(\phi, X)\right. \nonumber \\
	&\left. + \frac{h(\phi)}{2N}\left(\dot{\chi} - \beta^{i}\partial_{i}\chi\right)^{2} 
	- \frac{Nh(\phi)}{2}\gamma^{ij}\partial_{i}\chi\partial_{j}\chi \right].
	\label{eq:S_gene}
\end{align}

The constraint equations 
derived from the Lagrangian \eqref{eq:S_gene} are
\begin{align}
\rm{lapse}:\ 0 =& 
\frac{1}{2}f(\phi)\left(R^{(3)} - \frac{1}{N^{2}}(E^{ij}E_{ij} - E^{2})\right) 
+ \frac{f_{\phi}E}{N^{2}}(\dot{\phi}-\beta^{i}\partial_{i}\phi)  \nonumber \\
&- \gamma^{ij}\nabla^{(3)}_{i}\partial_{j}f(\phi)
+K(\phi, X) - \frac{K_{X}}{N^{2}}\left(\dot{\phi} - \beta^{i}\partial_{i}\phi\right)^{2}, 
\label{eq:const-gene1} \\
\rm{shift}:\ 0 =& 
\nabla^{(3)}_{i}\left[\frac{1}{N}f(\phi)(E^{i}_{j} - \delta^{i}_{j}E)\right] 
+ \frac{f_{\phi}E}{N}\partial_{j}\phi  \nonumber \\
&-\nabla^{(3)}_{j}\left[\frac{f_{\phi}}{N}(\dot{\phi}-\beta^{i}\partial_{i}\phi)\right]
- \frac{K_{X}}{N}(\dot{\phi}-\beta^{i}\partial_{i}\phi)\partial_{j}\phi,
\label{eq:const-gene2}
\end{align}
where again we neglect $\chi$-dependent terms here. 
We now expand the scalar field and metric around the background part as follows:
\begin{align}
	N &= 1+\alpha, \nonumber \\
	\beta^{i} &= \beta^{i}_{\rm{T}}, \ \ \partial_{i}\beta^{i}_{\rm{T}} = 0, \nonumber \\
	\gamma_{ij} &= e^{2\sigma + 2\zeta}(e^{h})_{ij}, \ \ \partial_{i}h_{ij} = h_{ii} = 0, \nonumber \\
	\phi &= \bar{\phi} + \delta\phi, 
\end{align}
and treat $\alpha$, $\beta_{\rm{T}}^{i}$, $\zeta$, $h$, $\delta\phi$ as perturbations. 
The background scale factor is given by $a(t)=e^{\sigma}$.
Here we solve the constraint equations \eqref{eq:const-gene1} and \eqref{eq:const-gene2} to first order in the perturbations.
In order to do this, notice that the $K_{X}$-dependent terms can be neglected since they are second order
in the perturbations. Moreover, we can take $K(\phi, X) = K(\phi, 0) \equiv V(\phi)$ since $X$ is a
second order quantity.
We again expand $f(\phi)$ to first order in the perturbations:
\begin{align}
	f(\phi) &\simeq f(\bar{\phi}) + f_{,\phi}(\bar{\phi})\delta\phi = M_{P}^{2}\left(1 + c_1\frac{\delta\phi}{M_{P}}\right),
\end{align}
where we set $f(\bar{\phi}) \equiv M_{P}^{2}$ and $f_{,\phi}(\bar{\phi}) \equiv c_1M_{P}$ in the second line. 
By noting these facts, it is straightforward to solve Eqs.~\eqref{eq:const-gene1} and \eqref{eq:const-gene2}. The solutions are
\begin{align}
	\alpha = \zeta &= -\frac{c_1}{2}\frac{\delta\phi}{M_{P}}, \nonumber \\
	\beta_{\rm{T}}^{i} &= 0,
\end{align}
where we used the background equations of motion \eqref{eq:bgeom-gene1} and \eqref{eq:bgeom-gene2}. Note that these solutions are
the same as those in Eq.~\eqref{eq:sol1}.
Then turning back to the Lagrangian \eqref{eq:S_gene}, 
the last term is expanded up to first order in the perturbations as
\begin{align}
	S &= \int d^4x\, a(t)^3\frac{1}{2}\left[ \left(d_1 \frac{\delta\phi}{M_{P}} + 3\zeta-\alpha\right) \dot\chi^2 
	- \left(d_1\frac{\delta\phi}{M_{P}}+\zeta+\alpha\right)
	\frac{(\partial_{i}\chi)^2}{a^2}  \right]\nonumber \\
	&= \int d^4x\, a(t)^3\frac{1}{2}\left(d_1- c_{1}\right)\frac{\delta\phi}{M_{P}} \left[ \dot\chi^2 - \frac{(\partial_{i}\chi)^2}{a^2} \right].
\end{align}
This coupling is totally the same as that of the minimal case. In particular, the coupling vanishes for $d_1=c_1$.

\section{Calculation in Another Gauge Condition}
For the check of our calculation in Sec. \ref{sec:fluc}, in this section we take a different gauge condition
from Eq.~\eqref{eq:gauge1}. We expand the fields as
\begin{align}
N &= 1+\alpha, \nonumber \\
\beta^{i} &= \partial_{i}\psi + \beta^{i}_{\rm{T}}, \ \ \partial_{i}\beta^{i}_{\rm{T}} = 0, \nonumber \\
\gamma_{ij} &= e^{2\sigma}(e^{h})_{ij}, \ \ \partial_{i}h_{ij} = h_{ii} = 0, \nonumber \\
\phi &= \bar{\phi} + \delta\phi, 
\label{eq:gauge2}
\end{align}
and treat $\alpha, \psi, \beta^{i}_{T}, h_{ij}, \delta\phi$ as perturbations. In the same way as before, 
the constraint equations \eqref{eq:constraints1} and \eqref{eq:constraints2} can be solved to obtain
\begin{align}
\alpha &= \frac{c_{1}}{2M_{P}H}e^{\sigma}\frac{d}{dt}\left(e^{-\sigma}\delta\phi \right), \nonumber \\
\psi &= -\frac{c_{1}}{2M_{P}H}e^{-2\sigma}\delta\phi, \nonumber \\
\beta^{i}_{T} &= 0.
\label{eq:sol2}
\end{align}
We consider $\phi(\partial h)^2$ couplings in this gauge. The relevant part  can be 
written as
\begin{align}
\left[\frac{1}{2}\sqrt{-g}f(\phi)R\right]_{\delta\phi hh} &=
\frac{c_{1}M_{P}^{2}}{8}\frac{\delta\phi}{M_{P}}
\left(e^{3\sigma}\dot{h}_{ij}\dot{h}_{ij} - e^{\sigma}\partial_{l}h_{ij}\partial_{l}h_{ij}\right) \nonumber \\
&-\frac{M_{P}^{2}}{8}\alpha\left(e^{3\sigma}\dot{h}_{ij}\dot{h}_{ij} + e^{\sigma}\partial_{l}h_{ij}\partial_{l}h_{ij}\right) \nonumber \\
&-\frac{M_{P}^{2}}{4}\partial_{l}\psi\partial_{l}\gamma_{ij}\dot{\gamma}_{ij}.
\end{align}
By substituting the solutions \eqref{eq:sol2} and doing some integration by parts, we finally obtain the 
following result:
\begin{align}
\left[\frac{1}{2}\sqrt{-g}f(\phi)R\right]_{\delta\phi hh} 
&= -\frac{c_{1}}{2H}\frac{\delta\phi}{M_{P}}\dot{h}_{ij}
\frac{\delta \mathcal{L}_{hh}}{\delta h_{ij}}\bigg|_{1\rm{st}},
\label{eq:stt2}
\end{align}
where $\delta \mathcal{L}_{hh}/\delta h_{ij}|_{1\rm{st}}$ is the variation of the kinetic term of the graviton 
$\mathcal{L}_{hh}$, which is given as
\begin{align}
\mathcal{L}_{hh} = \frac{M_{P}^{2}}{8}\left[e^{3\sigma}\dot{h}_{ij}\dot{h}_{ij} 
- e^{\sigma}\partial_{l}h_{ij}\partial_{l}h_{ij} \right].
\end{align}
The term remaining in Eq.~\eqref{eq:stt2} can be absorbed by the following field redefinition:
\begin{align}
\tilde{h}_{ij} = h_{ij} - \frac{c_{1}}{2H}\frac{\delta\phi}{M_{P}}\dot{h}_{ij}.
\label{eq:redef}
\end{align}
After such a field redefinition, there is no $\phi(\partial h)^2$. Therefore, a scalar
field cannot decay into gravitons in this gauge. This is consistent
with our previous result. In fact, the solutions in the present gauge \eqref{eq:sol2} and \eqref{eq:redef}
are related to those in the previous gauge \eqref{eq:sol1} by a gauge transformation.\footnote{
For gravitons, there are other second order contributions from the gauge transformation, which 
we neglect here just for simplicity. Such contributions arise, for example, from the field redefinitions
needed for other cubic terms; see Ref.~\cite{Maldacena:2002vr}.
} This completes the consistency check of our calculation.



\end{document}